\documentclass[useAMS,usegraphicx,usenatbib]{mn2e}
\usepackage{graphicx}
\usepackage{amsmath}
\usepackage{amssymb}
\usepackage{subfigure}
\usepackage{url}
\usepackage{multirow}

\newcommand{\s}{\rm\thinspace s}
\newcommand{\ks}{\rm\thinspace ks}

\newcommand{\Ms}{\rm\thinspace Ms}

\newcommand{\Hz}{\rm\thinspace Hz}

\newcommand{\Msun}{\hbox{$\rm\thinspace M_{\odot}$}}

\newcommand{\keV}{\rm\thinspace keV}

\newcommand{\degword}{\rm\thinspace deg}

\newcommand{\rg}{\rm\thinspace $r_\mathrm{g}$}

\title[The Origin of Lag Spectra]{The origin of the lag spectra observed in AGN: Reverberation and the propagation of X-ray source fluctuations}
\author[D. R. Wilkins \& A. C. Fabian]{D. R. Wilkins
  \thanks{E-mail: drw@ast.cam.ac.uk} 
and A. C. Fabian\\Institute of Astronomy, University of Cambridge, Madingley Road, Cambridge CB3 0HA}
\begin{document}

\date{Accepted 2012 December 10. Received 2012 December 7; in original form 2012 September 8}

\pagerange{\pageref{firstpage}--\pageref{lastpage}} \pubyear{2012}

\maketitle

\label{firstpage}

\begin{abstract}
The X-ray emission from active galactic nuclei (AGN) is highly variable. Measurements of time lags (characterised by lag spectra) between variability in the light curves in energy bands corresponding to directly observed continuum emission from the corona around the black hole and to X-rays reflected from the accretion disc adds a further dimension to studies of the structure and energetics of these systems. We seek to understand these measurements in terms of the physical parameters of the X-ray source (its location, extent, \textit{etc.}) through the calculation of theoretical lag spectra for a range of source parameters in general relativistic ray tracing simulations, combined with knowledge of the observed variability of the X-ray emission from AGN. Due to the proximity of the emission to the central black hole, Shapiro delays are important and the effects of general relativity should be considered when interpreting the lags as the light travel time between the source and reflector. We show that it is important to consider dilution of the lag by the contribution of both the primary and reflected spectral components to the observed energy bands rather than observing pure continuum and reflected emission, reducing the measured lag by up to 75 per cent compared to the `intrinsic' time lag due to light travel times. We find that the observed lag spectrum of the narrow line Seyfert 1 galaxy 1H\,0707-495 implies an X-ray source extending radially outwards to around 35\rg\ and at a height of around 2\rg\ above the plane of the accretion disc, consistent with the constraints obtained independently by considering the emissivity profile of the accretion disc. By investigating the influence of the propagation of X-ray luminosity fluctuations through the source region we find it is possible to reproduce the shape of the low frequency part of the lag spectrum (where the hard `primary' band lags behind the soft `reflected' band) as the effect of luminosity fluctuations originating in the centre of the X-ray source, close to the black hole, and propagating outwards.
\end{abstract}


\section{Introduction}
High resolution X-ray spectra from long observations of active galactic nuclei (AGN) have yielded an unprecedented amount of information about the structure of these systems, detecting emission from material right down to the innermost stable orbit and even the event horizon probing not only the mechanisms at work in AGN but also the strong gravity regime close to the black hole itself.

The observed X-ray spectra can be analysed in the context of emission from a primary X-ray source in a corona of energetic particles close to the central black hole, which inverse-Compton scatter the thermal photons emitted from the accretion disc up to X-ray energies. This emission is followed by reflection from an accretion disc \citep{george_fabian} which is geometrically thin in the equatorial plane and optically thick \citep{shaksun}. Detailed analysis of reflection features in the spectrum, notably the K$\alpha$ emission line of iron at 6.4\keV, has allowed constraints to be placed on the properties including the location and extent of the primary X-ray source \citep{wilkins_fabian_2011a, wilkins_fabian_2012}.

The X-ray emission from AGN is highly variable and recent studies of this variability have added a further dimension to the studies of these systems. The narrow line Seyfert 1 galaxy 1H\,0707-495 was the first AGN found to exhibit reverberation lags \citep{fabian+09}, where the variability in the spectral band dominated by reflection from the accretion disc is found to lag behind that in the spectral band corresponding to direct emission from the coronal X-ray source by tens to hundreds of seconds \citep{zoghbi+09}. Since then, reverberation lags have been discovered in a multitude of other AGN \citep{emmanoul+2011,demarco+2011,zoghbi+2011,zoghbi+2012,demarco+2012} and even in Galactic stellar mass X-ray binaries where the black hole mass and length scale invariance gives equivalent reverberation lags of the order milliseconds \citep{uttley+2011}.

\begin{figure}
	\centering
	\includegraphics[width=85mm]{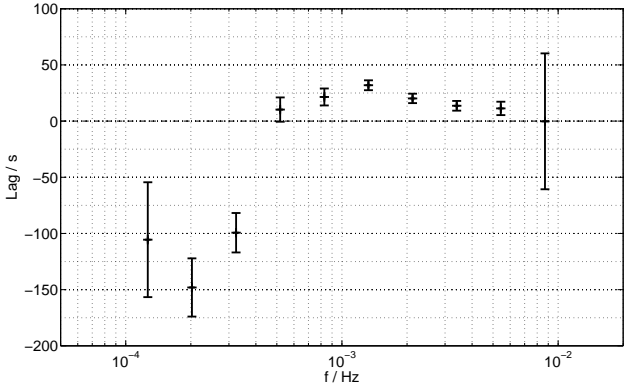}
	\caption{The lag spectrum from a 1\Ms\ lightcurve of 1H\,0707-495 from \citet{kara+12} showing the time lag for the variability at different Fourier frequencies between the hard (1.0-4.0\keV) band corresponding to the primary continuum emission and the soft (0.3-1.0\keV) band corresponding to reflection from the accretion disc. A positive lag indicates that variability in the soft (reflected) band lags behind that in the hard (primary) band.}
	\label{lagspec_1h0707.fig}
\end{figure}

Measuring these phase lags measures the light travel time between the source and reflector and allows us to probe scales as small as 10 light-seconds in AGN. The reverberation of X-rays from the accretion disc is characterised by the lag spectrum \citep{nowak+99} which shows the time lag of variability at different frequencies between the energy bands. The lag spectrum of 1H\,0707-495 was first obtained from a 500\ks\ lightcurve by \citet{zoghbi+09} and then in more detail following observations with XMM Newton totalling 1.3\Ms\ by \citet{kara+12}. Such a lag spectrum, from a 1\Ms\ lag spectrum of 1H\,0707-495 between the `hard' (1.0-4.0\keV) band dominated by the primary continuum and the `soft' (0.3-1.0\keV) band dominated by reflection from the accretion disc is shown in Fig. \ref{lagspec_1h0707.fig}.

At temporal frequencies of around $10^{-3}$\Hz, a lag of around 30\s\ can be seen between the arrival of the X-ray continuum and the reflection from the accretion disc. Na\"{i}vely, this can be converted into the distance of the reflecting material from the primary X-ray source, which taking the mass of the black hole in 1H\,0707-495 to be $2\times10^6\Msun$, \citep{zhou_wang}, gives a distance of 2\rg\ (a gravitational radius, 1\rg\ $=\frac{GM}{c^2}$). This simple calculation, however, takes into account neither the multiple paths that X-rays may follow from the source to the reflector nor the effects of general relativity which are expected to be significant in such close proximity to the black hole. It is assumed that the spectral energy bands correspond to detections of pure continuum and reflected emission, while in reality there will be contributions from both spectral components in each energy band. Finally, this analysis does not explain the full shape of the lag spectrum, rather it just looks at the longest time lag where the reflection follows the primary continuum.

The reverberation of of the iron K$\alpha$ emission line from the accretion disc of an AGN due to a localised flare of X-ray emission from the corona was first considered by \citep{fabian+89} and by \citet{stella-90} in the context of measuring the mass of the central black hole. \citet{gilfanov+2000} quantified the effect of reverberation in the Galactic black hole binary Cygnus-X1, computing an approximate solution neglecting relativistic effects while \citet{reynolds+99} accounted for the transport of the observed X-rays in general relativity to compute `2D transfer functions' giving the observed flux as a function of observed photon energy as time elapses from a single, localised, X-ray flare. \citet{young_reynolds} related this work to observable signatures that may be detected by future generation X-ray missions in the context of comparing observed reverberation of X-ray flares with a library of computed transfer functions to determine the mass of the black hole as well as the location of the X-ray flare.

We here present an analysis of X-ray reverberation applicable to the current state-of-the-art observations with XMM Newton. A systematic analysis of lag spectra computed theoretically for reflection of X-rays originating from a variety of X-ray sources of different sizes and geometries and in different locations allows us to understand the observed form of the lag spectra in AGN and to ask how much can be learned from these measurements about the energetics and geometry of these systems. While applicable generally to X-ray sources exhibiting a reverberation lag, this study is motivated by the observed lag spectrum of 1H\,0707-495. As well as the location and geometry of the X-ray source, we consider observational constraints in separating the detection of the primary continuum and the reflected X-rays from the accretion disc and we consider how variations in luminosity may propagate through the X-ray source region.

\section{Lag Spectra}
Following \citet{nowak+99}, time lags between spectral components (with their own light curves) are characterised by the lag spectrum, showing the phase (and hence) time lag between the Fourier frequency components that make up the light curves. The light curve, $F(t)$, is considered to be the sum over components of all frequencies:
\[ F(t) = \int \tilde{F}(\omega)e^{i\omega t}\,d\omega \]

The amplitude and phase of each frequency component is given by the Fourier transform of the light curve and can be written $\tilde{F}(\omega)=\left|\tilde{F}(\omega)\right|e^{i\varphi}$ and is computed by
\[ \tilde{F}(\omega) = \int F(t)e^{-i\omega t}\,dt \]
The phase lag between two spectral components, say the hard and soft light curves, $H(t)$ and $S(t)$ respectively, can be found by considering the complex form of their Fourier transforms $\tilde{H}(\omega)=\left|\tilde{H}(\omega)\right|e^{i\varphi}$ and $\tilde{S}(\omega)=\left|\tilde{S}(\omega)\right|e^{i\theta}$ and computing the cross spectrum
\begin{equation}
	\label{crossspec.equ}
	\tilde{C}(\omega) = \tilde{S}^*(\omega) \tilde{H}(\omega) = \left|\tilde{S}(\omega)\right| \left|\tilde{H}(\omega)\right| e^{i(\varphi - \theta)}
\end{equation}
The argument of which gives the time lag, $\tau$, since $\varphi = \omega t$ (and converting from angular to linear frequency, $f$)
\begin{equation}
	\label{lag.equ}
	\tau(f) = \frac{1}{2\pi f}\arg\left(\tilde{C}(f)\right)
\end{equation}
Following this calculation (and sign convention), a positive time lag indicates that the variability in the soft energy band, $S(t)$, is lagging behind that in the hard energy band, $H(t)$, as would be expected if the hard band is dominated by an X-ray continuum which is reflecting off of the accretion disc and this reflected radiation dominates the soft band.

We here issue a warning to the reader that this definition is the opposite sense of the time lag as adopted by \citet{zoghbi+09} and \citet{nowak+99} who define the soft band lagging behind the hard band to be the negative lag, though defining this as a positive lag is more instructive when considering the features of lag spectra arising from reverberation of X-ray variability from a reflecting accretion disc.

Detection of a time lag between spectral components requires the light curves to be coherent (\textit{i.e.} there is a constant phase lag between the Fourier components over a given range in the two light curves, rather than a random one), although \citet{kara+12} argue that total coherence is not required as uncorrelated (incoherent) variability will not contribute to the calculated lag as this part will average to zero.

\begin{figure}
	\centering
	\includegraphics[width=85mm]{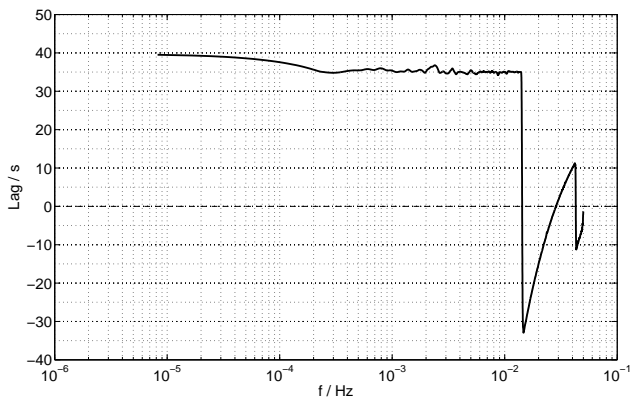}
	\caption{Lag spectrum between a primary light curve (taken from the light curve of 1H\,0707-495 in the 1-4\keV\ band) and the same light curve delayed by 30\s. The lag spectrum shows a constant lag of approximately the input 30\s\ across all frequency components until phase-wrapping occurs when the lag corresponds to a half-wave shift in phase at $f = \frac{1}{2\tau}$. The drift to slightly a longer lag at the lowest frequencies is due to the error in the Fourier transform of the observed light curve.}
	\label{lagspec_shift.fig}
\end{figure}

To explore the general features of reverberation lags, the lag spectrum was computed for the simplest case of a `reflected' light curve exactly following the primary light curve but delayed in arrival by 30\s\ (\textit{i.e.} shifting the time axis by 30\s). Such a lag spectrum, using the light curve of 1H\,0707-495 in the 1-4\keV\ band as the input is shown in Fig. \ref{lagspec_shift.fig}.

The lag spectrum shows a constant lag of approximately the input 30\s\ across all frequency components until the lag corresponds to a half-wave shift in phase at $f = \frac{1}{2\tau}$ (give or take the error introduced by the numerical, discrete Fourier transform of the real light curve and the temporal binning of the light curve, the combination of which causes the drift to a slightly longer lag at the lowest frequencies). Once $f = \frac{1}{2\tau}$, the waveform could have been shifted either forwards or backwards by half a wave and since the phase angle from which the lag is calculated is defined to be in the range $-\pi < \varphi < \pi$, phase wrapping occurs. At higher frequencies, the phase continually wrap causing the lag to oscillate around zero until it decays away. It is not possible to measure lags longer than a shift of half a wavelength.

\section{Simulating Lag Spectra for X-ray Reflection}
\subsection{The Transfer Function}
The time lag between the observation of a change in the primary X-ray continuum and the corresponding change in the reflected flux from the accretion disc is characterised by the \textit{transfer function}, $T(t)$. The transfer function is the response seen from the accretion disc to an instantaneous flash of light from the primary X-ray source.

The transfer function is calculated by general relativistic ray tracing simulations in two stages. Initially, rays are transferred from the source to the accretion disc, in the equatorial plane of the Kerr geometry around the central black hole with dimensionless spin parameter, $a$, as described in \citet{wilkins_fabian_2012}.

Initially, the X-ray source is assumed to be an isotropic point source, emitting equal power into equal solid angle in its own instantaneous rest frame. The photon's initial 4-momentum is transformed from the source's frame, a locally flat spacetime described by a tetrad of orthonormal basis vectors, to the global Boyer-Lindquist co-ordinate system to calculate the constants of motion of the ray, $k$, $h$ and $Q$. The ray is then traced by numerical integration of the null geodesic equations (the equations of motion of a photon) in the Kerr spacetime. Taking $c=1$ and $\frac{GM}{c^2}=1$ to work in natrual units of $r_\mathrm{g} = \frac{GM}{c^2}$, the geodesic equations can be written:
\begin{eqnarray*}
	\label{tdot.equ}	
	\dot{t} &=& \frac{
		\left[(r^2 + a^2\cos^2\theta)(r^2+a^2) + 2a^2 r\sin^2\theta\right]k - 2arh
	}
	{ 
		r^2\left(1+\frac{a^2\cos^2\theta}{r^2} - \frac{2}{r}\right)\left(r^2+a^2\right) + 2a^2r\sin^2\theta
	} \\
	\label{phidot.equ}
	\dot{\varphi} &=& \frac{
		2ar k \sin^2\theta + (r^2+a^2\cos^2\theta - 2r)h
	}
	{
		(r^2+a^2)(r^2+a^2\cos^2\theta-2r)\sin^2\theta + 2a^2 r\sin^4\theta
	}
\\
	\label{thetadot.equ}
	\dot{\theta^2} &=& \frac{Q + (ka\cos\theta - h\cot\theta)(ka\cos\theta + h\cot\theta)}{\rho^4}
\\
	\label{rdot.equ}
	\dot{r}^2 &=& \frac{\Delta}{\rho^2}\left[ k\dot{t} - h\dot{\varphi} - \rho^2\dot\theta^2  \right]
\end{eqnarray*}
Where the dots denote differentiation with respect to some affine parameter, $\sigma$.

Spatially extended X-ray sources may be explored through Monte Carlo ray tracing simulations. Rays are started at random locations (with a uniform probability density function such that all locations are equally likely) within an allowed cylindrical source region (defined by a lower and upper height from the disc plane as well as an inner and outer radius) and are assigned random initial direction cosines $\cos\alpha$ and $\beta$, again with uniform probability density. This will simulate the effect of an X-ray source of finite spatial extent that is optically thin to the X-rays it emits. Each local region of the source is taken to be co-rotating with the element of the accretion disc in a relativistic Keplerian orbit at the same radius. This model does, however, assume the surface brightness over the entire extent of the source varies simultaneously, which may be unphysical if the source region extends across multiple gravitational radii, as the light crossing time of the source is no longer negligible compared to the light travel time from the source to the reflector, however this model may be used to approximate the behaviour of an extended source and we will revisit the propagation of fluctuations in \S\ref{prop.sec}.

When the rays reach the accretion disc, their arrival time as measured by an observer at infinity (the $t$ co-ordinate) as well as position ($r,\varphi$) are recorded and they are sorted into radial and azimuthal bins on the accretion disc.

The travel time of photons from each radial and azimuthal bin on the accretion disc as well as their energy measured by an observer at infinity is determined by tracing parallel rays, travelling perpendicular to a flat image plane a large distance from the black hole, representing the area of the sky imaged by the X-ray telescope. Rays are started on a regular grid (the `image plane') travelling perpendicular to this plane plane, a distance 10000\rg\ from the black hole (to be certain the spacetime here is flat), inclined to the rotation axis of the black hole at the angle at which the telescope is observing the system, henceforth taken to be 53\degword, the measured inclination of 1H\,0707-495. The image plane is centred on the black hole and spans the area of interest (\textit{i.e.} the accretion disc, assumed to have a radius of 400\rg).

Due to the symmetry of the Kerr metric, propagating rays backwards in time (starting at the image plane to work out where they originated from on the accretion disc) is equivalent to propagating forwards in time, having reversed the direction of spin of the black hole. Propagating the rays from the image plane with the spin of the black hole reversed until they reach the equatorial plane will therefore give the position on the accretion disc ($r$ and $\varphi$ co-ordinates) from which the ray originated as well as the travel time of the ray from the disc to the telescope as measured by the observer at infinity ($t$ co-ordinate). The rays are sorted into the radial and azimuthal bins upon the accretion disc and the average travel time of a ray is calculated for each bin. The number of rays landing in each bin is counted, which gives the area of that bin projected onto the image plane by general relativistic light bending and will determine the flux received from each part of the accretion disc.

The energy and arrival rate of the photons arriving along a given ray measured by the observer at infinity is computed by the projection of the photon 4-momentum to the observers' timelike axes (which are their 4-velocities), though before doing so, since the photon under consideration is propagating backwards, the spacelike components of the 4-momentum are reversed (so that the photon is travelling in the correct direction with respect to the orbiting material in the accretion disc and the correct Doppler shift is computed).
\begin{equation}
	\label{redshift.equ}
 g^{-1} \equiv \frac{\nu_\mathrm{O}}{\nu_\mathrm{E}} = 
 \frac{\mathbf{v}_\mathrm{O}\cdot\mathbf{p}(\mathrm{O})}{\mathbf{v}_\mathrm{E}\cdot\mathbf{p}(\mathrm{E})} = \frac{g_{\mu\nu}v_O^\mu p^\nu(O)}{g_{\rho\sigma}v_E^\rho p^\sigma(E)}
\end{equation}
The average energy shift for each bin on the accretion disc is computed from all the rays originating from it. It is assumed that the primary X-ray source is emitting photons of all energies and that this will excite emission lines and other spectral features from the disc. The energy shift of the observed photons is assumed to arise only from the transport of rays from the accretion disc to the observer, taking into account special relativistic Doppler shifts as well as gravitational redshifts. Material in the accretion disc is assumed to be in a (relativistic) prograde Keplerian orbit in the equatorial plane around the black hole.

When a photon from the primary X-ray source lands in a specific bin on the accretion disc, the time from the disc to the observer as well as the energy of the photons measured at infinity is looked up from the values calculated for that bin from the rays traced from the image plane to the disc. The total arrival time of the ray from the primary X-ray source, via the disc, to the observer is the sum of the two computed times.

Finally, the transfer function is computed by counting the number of rays that arrive at the observer as a function of time, $N(t,r,\varphi,E)$, integrating over all energies and positions on the disc. Assuming the X-ray source emits photons at an equal rate along each ray in its own rest frame (\textit{i.e.} it is isotropic), the arrival rate of photons along each ray measured by the observer will vary according to the redshift, $\frac{\nu_\mathrm{O}}{\nu_\mathrm{E}}\equiv g^{-1}$, which is the product of the redshifts from the source to the disc ($g^{-1}_\mathrm{SD}$) and from the disc to the observer ($g^{-1}_\mathrm{DO}$).
\begin{equation}
	T(t) = \int N(t,r,\varphi,E)\thinspace g^{-1}_\mathrm{SD}g^{-1}_\mathrm{DO}\thinspace rdrd\varphi dE
\end{equation}

The ray tracing code to compute these transfer functions was implemented to run rapidly on graphics processing units (GPUs) using the \textsc{nvidia cuda} framework, allowing the paths of hundreds of rays to be computed in parallel (for more details of the implementation of general relativistic ray tracing codes on GPUs, see \citealt{wilkins_fabian_2012}).

\subsection{The Arrival Time of the Primary Continuum}
The ray tracing simulations described above give the photon arrival rate as a function of time since the initial flash of light left the primary X-ray source. However, when measuring reverberation in observations of AGN, it is the time lag between the arrival of the primary continuum and the reflected rays that is measured. One must therefore know the arrival time of the primary X-ray emission at the observer to define the zero in the time series of the transfer function.

For an infinitesimal point source, exactly one ray emitted from the source will end up at any given observer at infinity. Therefore the arrival time of the primary X-ray continuum may be computed by tracing one ray from the source to the observer. Table \ref{direct_ray.tab} shows the direct ray travel times (as measured by an observer at infinity) from point sources at varying heights above the black hole to an observer at a distance of 10000\rg\ from the black hole inclined at 53\degword\ to the rotation axis. A source above the plane of the accretion disc should be closer to the observer than the black hole and classically, in natural units, light takes $1\frac{GM}{c^3}$ to travel 1\rg. These direct rays taking longer than 10000\thinspace $\frac{GM}{c^3}$ to reach the observer illustrates the Shapiro delay here where the passage of light is slowed down as it travels in the proximity of the black hole \citep{shapiro}.

\begin{table}
\centering
\caption{Travel time (in natural units) as measured by an observer at infinity for a ray propagating from point sources at varying heights above the black hole on the rotation axis to an observer 10000\rg\ away inclined at an angle of 53\degword\ to the rotation axis. The Shapiro delay is characterised by the difference between the travel time of the ray in general relativity and what it would be in classical, Euclidean space.}
\begin{tabular}{ccc}
  	\hline
   	\textbf{Source Height} & \textbf{Ray Travel Time} & \textbf{Delay}\\
	\hline
	1.235\rg & 10029\thinspace $GM/c^3$ & 30\thinspace $GM/c^3$\\
	2\rg & 10020\thinspace $GM/c^3$ & 21\thinspace $GM/c^3$\\
	5\rg & 10014\thinspace $GM/c^3$ & 17\thinspace $GM/c^3$\\
	10\rg & 10008\thinspace $GM/c^3$ & 14\thinspace $GM/c^3$\\
	20\rg & 10001\thinspace $GM/c^3$ & 13\thinspace $GM/c^3$\\
	\hline
\end{tabular}
\label{direct_ray.tab}
\end{table}

\subsection{The Reflected Light Curve and Lag Spectrum}
The transfer function describes the response seen in the reflection from the accretion disc due to an instantaneous flash of X-rays from the primary source. Hence, if the primary X-ray source, observed in a `hard' band for the X-ray spectrum is varying as described by some light curve $H(t)$, the response in the reflection seen in a `soft' band, $S(t)$, is given by the convolution of the primary light curve with the transfer function.
\begin{equation}
	S(t) = H(t) \otimes T(t) = \int H(t') T(t - t')\,dt'
\end{equation}
Ray tracing in the simulations is carried out in natural units, measuring time in units of $\frac{GM}{c^3}$. The time series of the transfer function is trivially converted to seconds for the mass of the black hole of interest.

In order to generate a realistic lag spectrum with the power spectral density (PSD) observed in the primary continuum of an AGN, the primary, hard light curve is taken to be the observed light curve of the narrow line Seyfert 1 galaxy, 1H\,0707-495 in the 1-4\keV\ band in 10\s\ bins. Using the observed light curve of 1H\,0707-495 also means that lag spectra are simulated with a realistic time resolution that is achievable with the current generation of X-ray telescopes. The mass of the black hole is taken to be $2\times10^6\Msun$, as quoted in the literature \citep[see \textit{e.g.}][]{zhou_wang}.

The primary light curve is convolved with the transfer function obtained from the ray tracing simulations in order to obtain the light curve of the reflected component. The Fourier transforms of both the primary and reflected light curves were then computed. The cross and lag spectra are calculated using Equations \ref{crossspec.equ} and \ref{lag.equ}. The convolution, Fourier transforms (using the Fast Fourier Transform, FFT algorithm) and lag calculations were performed in \textsc{matlab}.

\subsection{Extended Sources}

For an extended source, a range of rays following different paths will be observed by the telescope as the primary continuum. A transfer function (photon arrival rate vs. time) for an extended source region may be calculated in exactly the same way as the rays are traced from the observer to the accretion disc. If the source is said to extend radially at some height above the accretion disc, rays are traced from the image plane until they reach this plane and their time is measured if they intercept the region of the plane spanning the source.

In order to compute the lag spectrum arising from the reflection of X-rays originating from an extended source, the observed light curve is taken to be the `intrinsic' variability of the X-ray source. This is permissible since the transfer functions do not impart any power on the variability spectrum, they just give a time lag and this will mean the simulated variability has a physically realistic power spectrum. Simulated light curves that would be observed for the primary (hard) and reflected (soft) spectral components, $H(t)$ and $S(t)$ respectively, are computed by convolving the light curve with the calculated transfer functions for each component.
\begin{align}
	\label{hconv.equ}
	H(t) &= H_\mathrm{obs}(t) \otimes T_\mathrm{src}(t) \\
	\label{sconv.equ}
	S(t) &= H_\mathrm{obs}(t) \otimes T_\mathrm{ref}(t)
\end{align}
The lag spectrum between the two spectral components is calculated as above.

\section{Theoretical Lag Spectra}

\subsection{Lag Spectra for Point Sources}
\begin{figure*}
\centering
\begin{minipage}{170mm}
\subfigure[]{
\includegraphics[width=85mm]{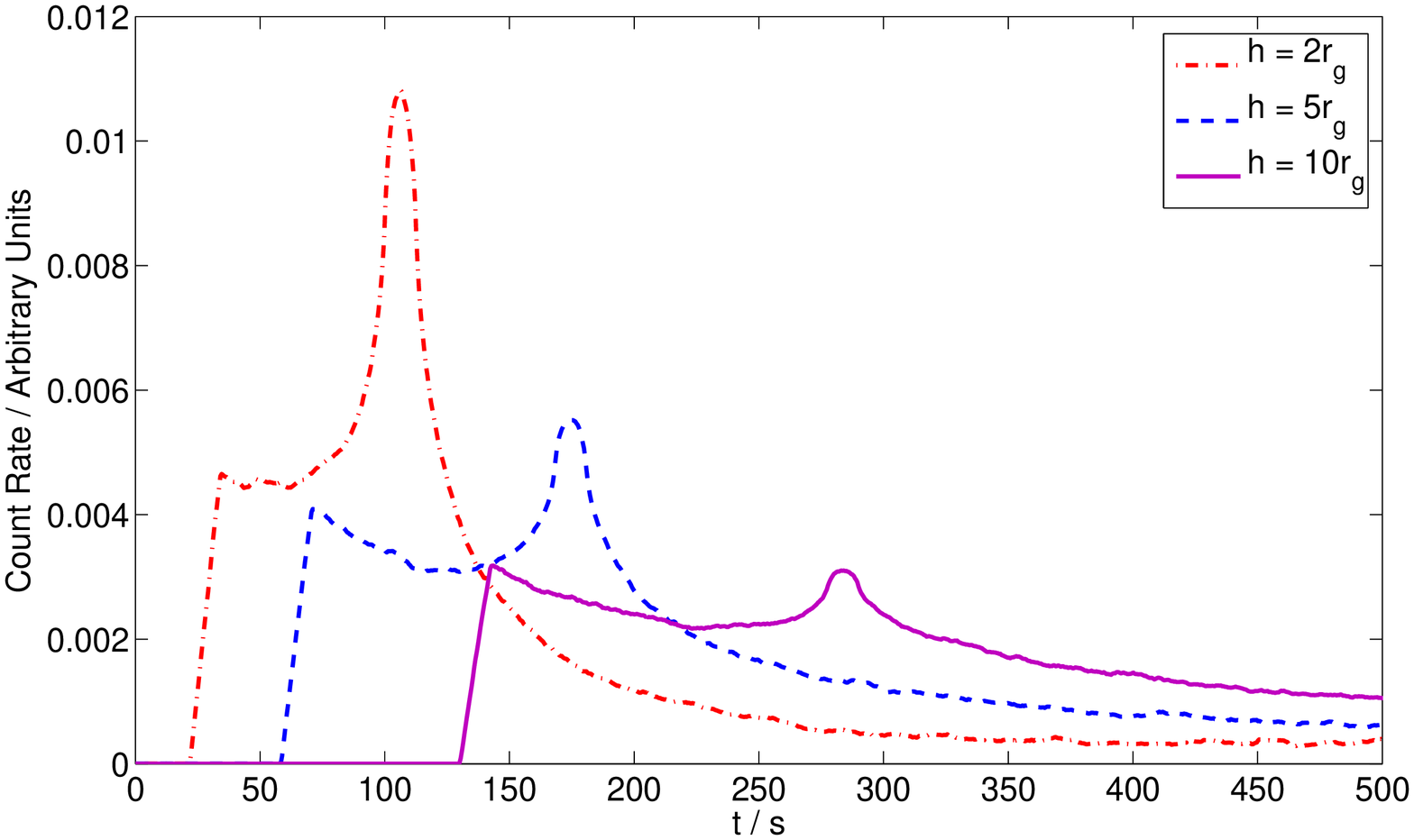}
\label{lags_ax.fig:trf}
}
\subfigure[]{
\includegraphics[width=85mm]{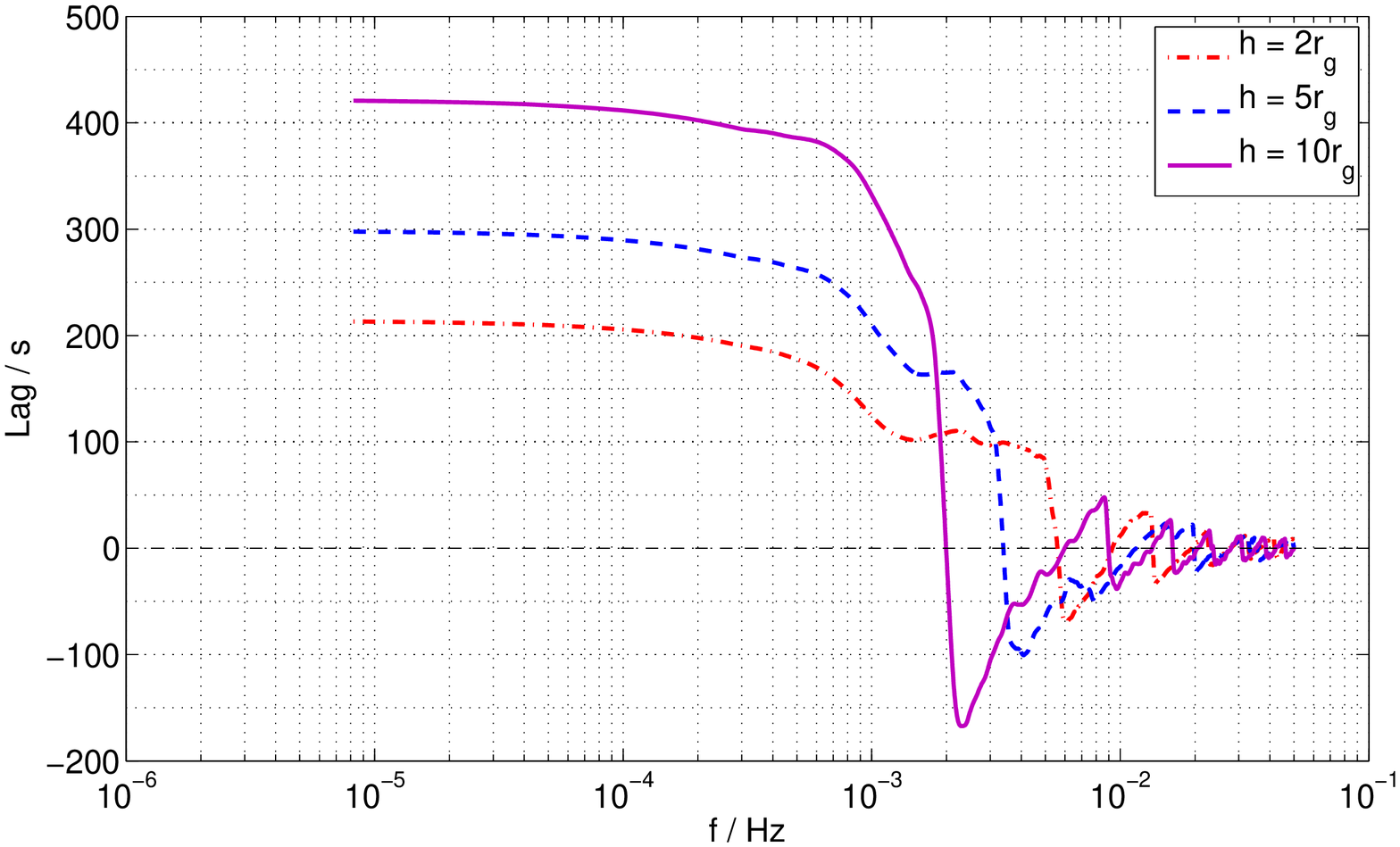}
\label{lags_ax.fig:spec}
}
\caption[]{\subref{lags_ax.fig:trf} Transfer functions and \subref{lags_ax.fig:spec} lag spectra for reflection due to isotropic point sources at varying heights, stationary on the rotation axis above the black hole, mass $2\times10^6\Msun$. The accretion disc is observed at an inclination angle of 53\degword. A positive time lag indicates that variability in the reflected X-rays lag behind variability in the primary emission from the source.}
\label{lags_ax.fig}
\end{minipage}
\end{figure*}

Initially, the transfer functions and lag spectra were computed for reflection from isotropic point sources located at varying heights on the rotation axis above the black hole, as shown in Fig. \ref{lags_ax.fig}.

A striking feature of the transfer functions is the peak in X-rays after the initial arrival of the reflection from the accretion disc. This corresponds to the `re-emergence' of the redshifted, gravitationally lensed (and thus magnified) reflection from the far side of the accretion disc, travelling round the black hole to the observer as was noted by \citet{reynolds+99}.

Turning to the resulting lag spectra, a constant lag is observed at low frequencies, corresponding to the average time lag of the transfer function
\begin{equation}
	\label{avglag.equ}	
	\bar{t} = \frac{1}{\int T(t)\,dt} \int t\,T(t)\,dt
\end{equation}

Towards higher frequencies, the contribution to the lag from the extended tail of the transfer function decreases as phase wrapping causes these contributions to the lag to average to zero. This decreases the average lag time at higher frequencies, until the peak in the transfer function is left to dominate, causing the lag spectrum to flatten off before the jump down where phase wrapping occurs for all components of the lag.

The transfer functions roughly follow the form of those computed by \citet{gilfanov+2000} who calculated the transfer functions for point sources in Euclidean spacetime, taking into account no relativistic effects and only for accretion discs extending inward to 10\rg. Here, the fully relativistic transfer functions are computed for accretion discs extending all the way down to the innermost stable orbit for a maximally spinning black hole. As in the classical case, the transfer functions fall of as $t^{-2}$ at long times, though the relativistic effects cause the initial peak in the transfer function to be steeper. Extending the accretion disc in to the innermost stable orbit means the flattened part visible at early times in the transfer functions of \citet{gilfanov+2000} are not seen (consistent with their trend as the inner radius of the disc is decreased).

Considering, for illustration, the case of the point source at a height of 5\rg, the mean arrival time of rays in the transfer function is around 290\s\ after the arrival of the primary continuum from the point source. This time can be compared with travel times computed for the equivalent source and reflector in flat, Euclidean spacetime. For the case of a point source at 5\rg, the primary continuum takes 168\s\ longer to reach the observer accounting for relativistic effects than it would in Euclidean spacetime and computing the transfer function using Euclidean ray travel times (but keeping the relativistic formulation for the projection of areas on the accretion disc to the observer and the emissivity profile of the disc), the average arrival time is delayed by 173\s. Since the lag is measured between the arrival of the continuum and the reflection, relativistic effects on the light travel time increase the measured lag by around 4\s\ (the `Shapiro delay'). When directly equating the lag to the light travel time from the source to the reflector, this means the source appears 0.5\rg\ further from the accretion disc than it really is.

\subsection{Lag Spectra for Extended Sources}
Whilst it is instructive to consider the lag spectra arising from point sources in general relativity, observations of the relativistically broadened iron K line profile in 1H\,0707-495 suggest that the X-ray source is radially extended out to around 35\rg\ \citep{wilkins_fabian_2011a,wilkins_fabian_2012}. It is therefore important to understand the form of lag spectra arising from extended sources.

While the following analysis is generally applicable to X-ray reverberation from spatially extended sources, we are motivated by the observed lag spectrum of 1H\,0707-495 and illustrate the effects with source geometries similar to that suggested by the emissivity profile observed in this object. Further more, we consider a source region whose luminosity is constant throughout its extent. While in reality the luminosity is likely to vary across the source, perhaps being concentrated in the central regions, we here elect to minimise the number of free parameters such that we may gain insight into the physical effects. When assuming a constant source luminosity, the modelled source extents will represent the bulk of the emitting material.

\begin{figure*}
\centering
\begin{minipage}{170mm}
\subfigure[]{
\includegraphics[width=85mm]{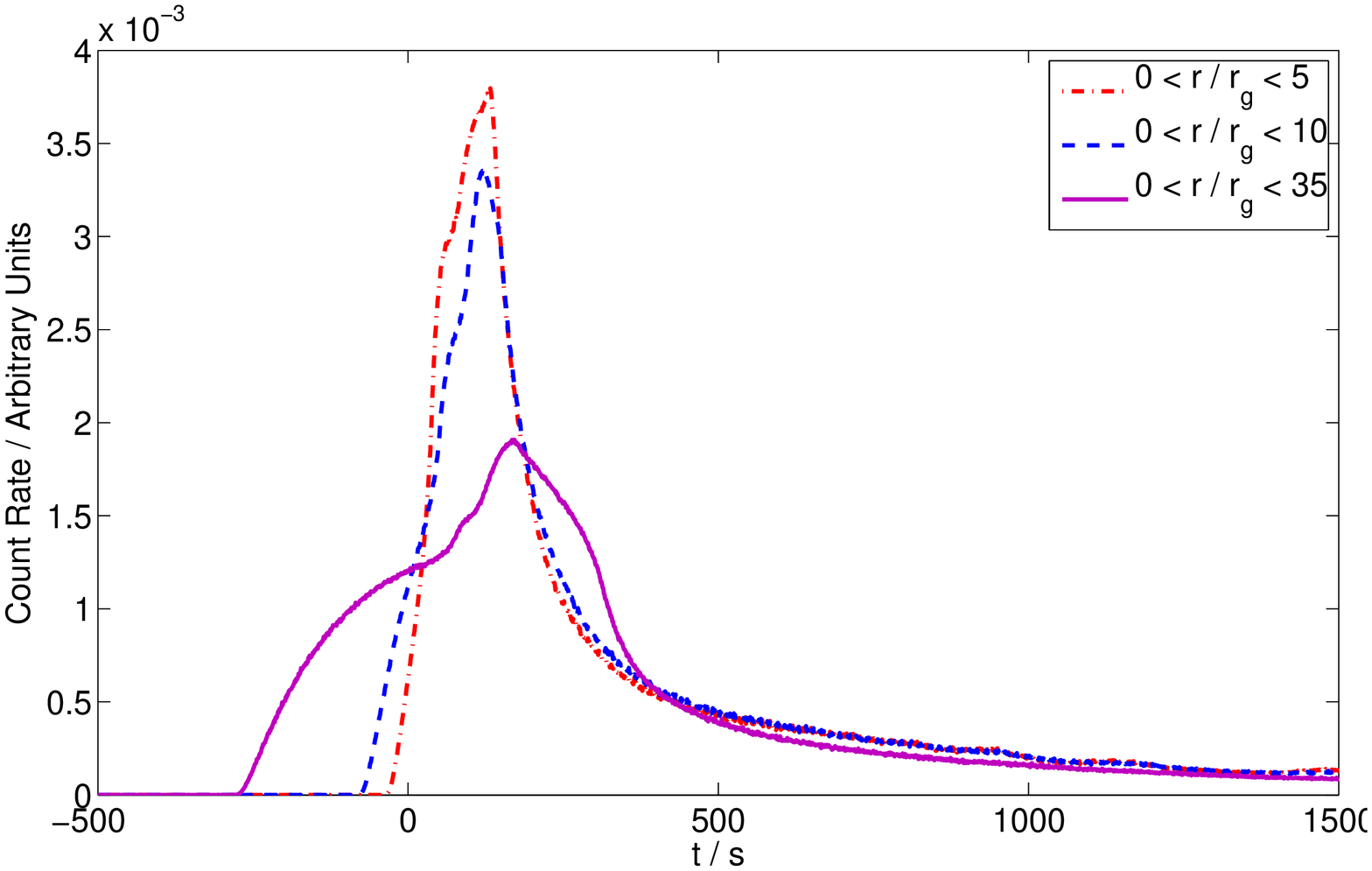}
\label{lags_radius.fig:trf}
}
\subfigure[]{
\includegraphics[width=85mm]{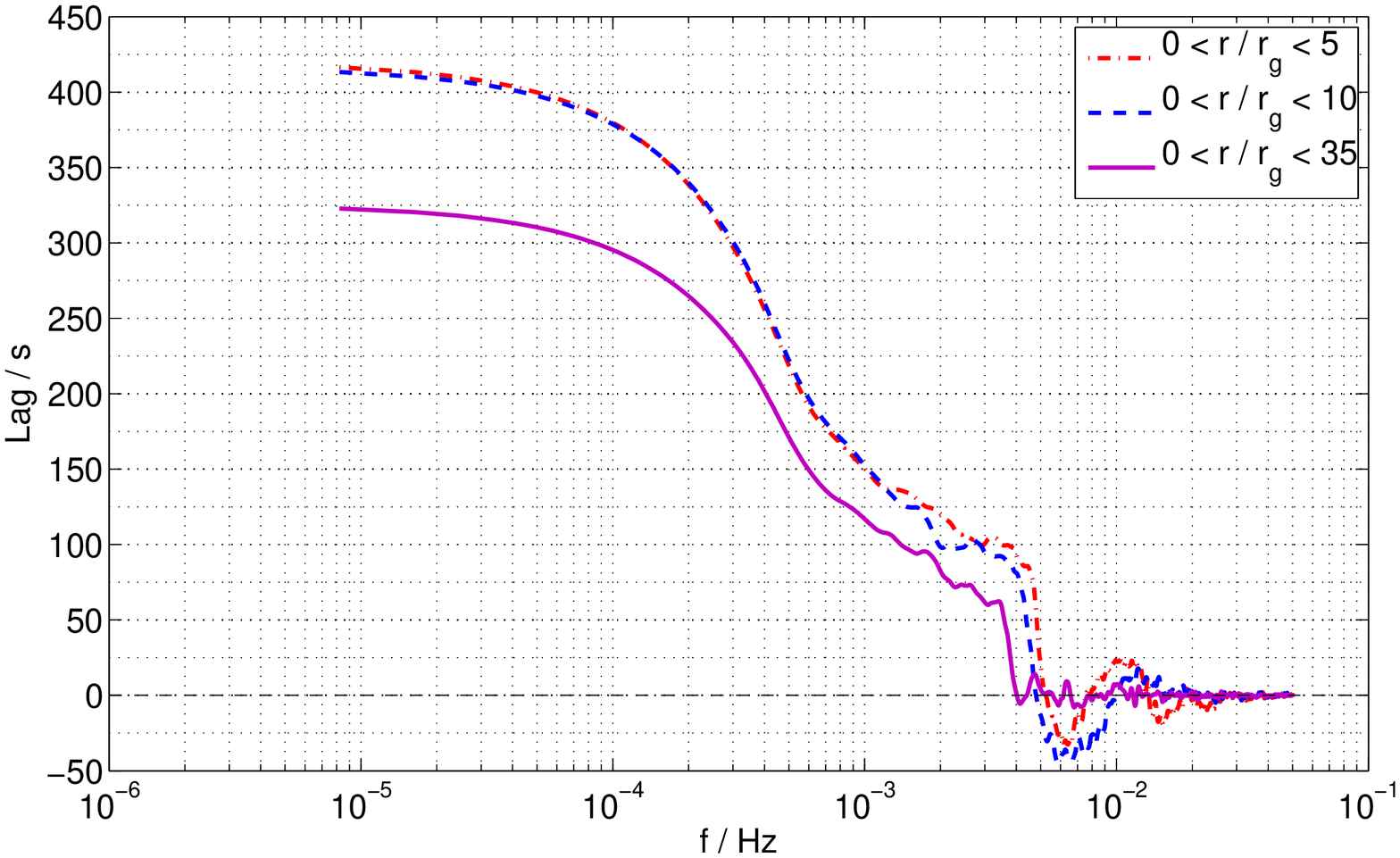}
\label{lags_radius.fig:spec}
}
\caption[]{\subref{lags_radius.fig:trf} Transfer functions and \subref{lags_radius.fig:spec} lag spectra for the instantaneous illumination of X-ray sources extended vertically between 2\rg\ and 2.5\rg\ above the plane of the accretion disc and with different radial extents outward from the rotation axis.}
\label{lags_radius.fig}
\end{minipage}
\end{figure*}

The transfer functions and lag spectra for reflection of X-rays from source regions of constant height and vertical extent but increasing in radial extent outward from the rotation axis are shown in Fig. \ref{lags_radius.fig}. The most notable feature of the transfer function for extended sources is the extended tail to long reverberation times since from an extended X-ray source, there are many possible light paths for the primary X-rays to reach the accretion disc. 

In addition to the extended tail of the transfer function, some reflected rays (those originating from the edge of the X-ray source closest to the observer reflecting off the part of the disc immediately below) may reach the telescope before the mean arrival time of the continuum, meaning the transfer function begins before the time $t=0$. As for the case of a point source, the reflected X-rays are seen to rise after their initial arrival owing to the re-emergence of X-rays from the far side of the accretion disc, lensed around the black hole, though this is less pronounced due to the increased range of ray paths now possible with an extended source.

Turning to the lag spectrum in Fig. \ref{lags_radius.fig:spec}, a constant lag is once again observed at low frequencies corresponding to the mean arrival time of reflected rays over the transfer function, though with an extended X-ray source, since more of the flux is in the extended tail of the transfer function, there is a much more pronounced decay of the lag measured in the higher frequency components. For the most extended sources, there is almost a continuous decay of the measured lag to zero in the high frequency components with no phase-wrapping observed. The extremes of the source being further from the black hole reduces the shortest possible light travel time to the observer since further from the black hole, the time delay due to spacetime curvature (the Shapiro delay) is reduced. This means that for the most radially extended sources, the observed lag is less than that measured for the sources more confined to the region around the central black hole.

\begin{figure*}
\centering
\begin{minipage}{170mm}
\subfigure[]{
\includegraphics[width=85mm]{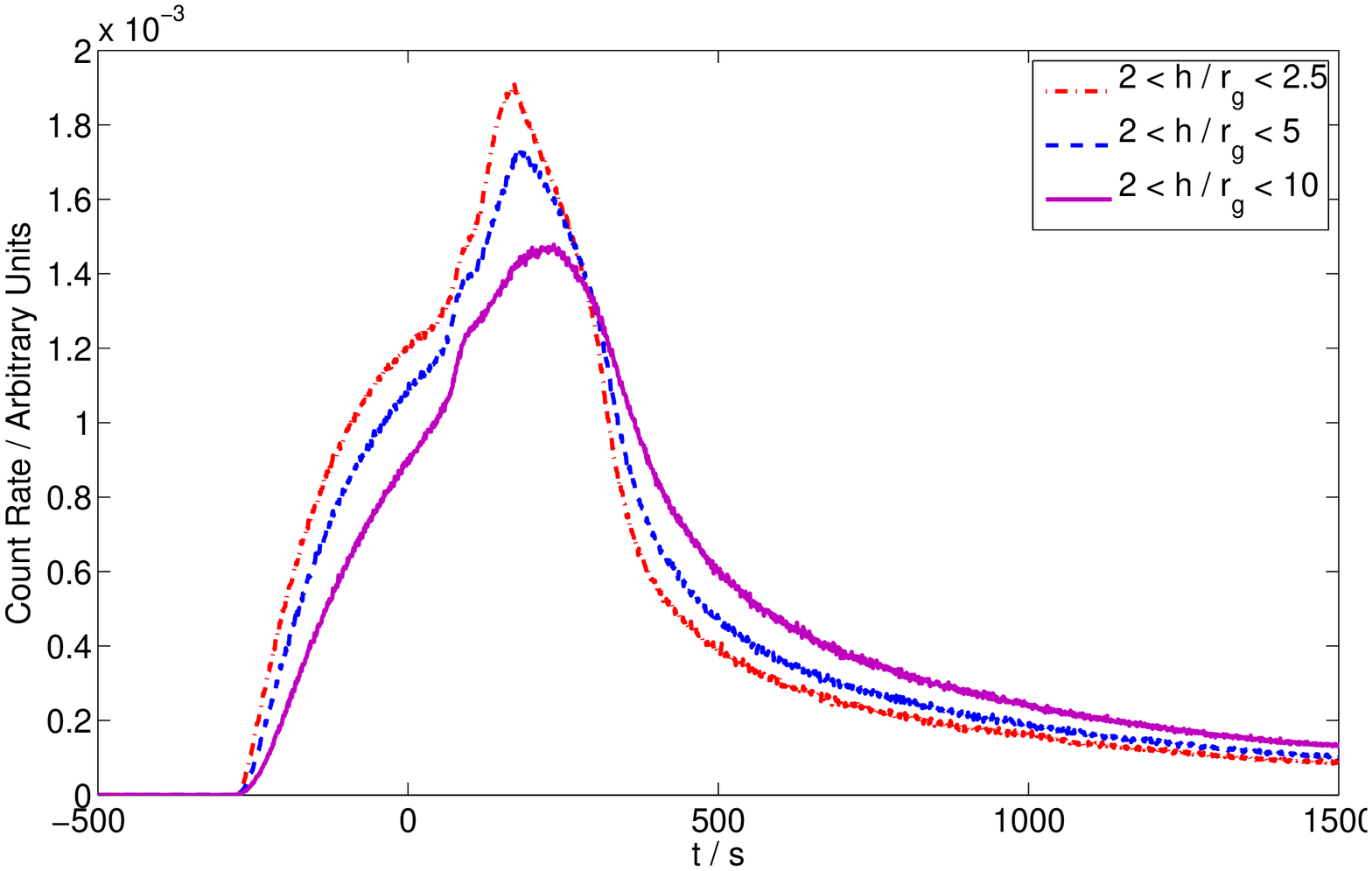}
\label{lags_height.fig:trf}
}
\subfigure[]{
\includegraphics[width=85mm]{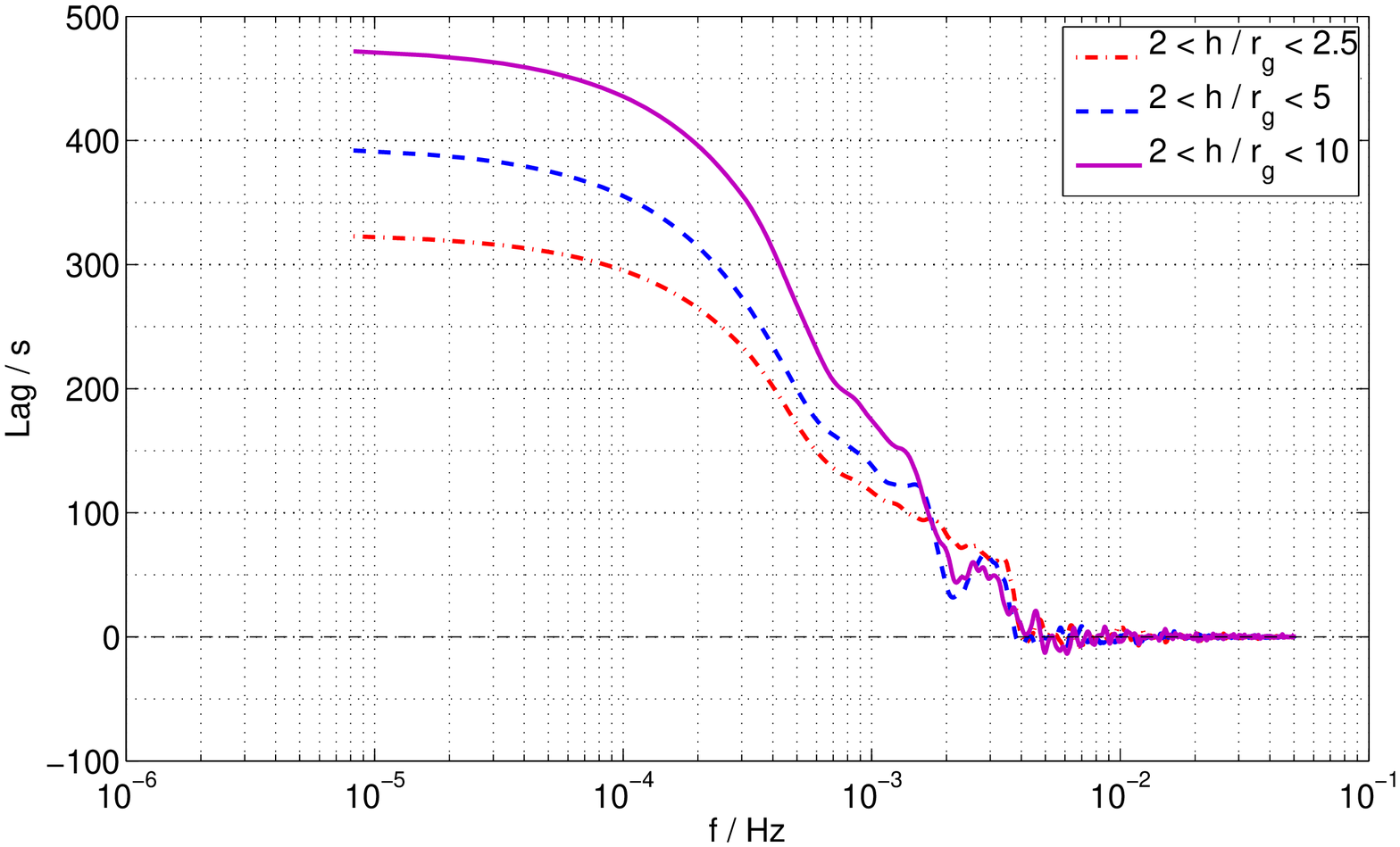}
\label{lags_height.fig:spec}
}
\caption[]{\subref{lags_height.fig:trf} Transfer functions and \subref{lags_height.fig:spec} lag spectra for X-ray sources extended vertically from 2\rg\ to varying heights above the plane of the accretion disc. In each case, the X-ray source is extended radially out to 35\rg.}
\label{lags_height.fig}
\end{minipage}
\end{figure*}

Increasing the vertical extent of the source region (Fig. \ref{lags_height.fig}) again increases the range of possible ray paths from the source to the disc, giving the transfer function an extended tail and causing the measured lag to decay away more for the higher frequency components of the variability. Increasing the vertical extent of the source above a fixed bottom also moves the mean source location away from the accretion disc, increasing the mean lag time measured for the lowest frequency components.

\section{Dilution of Spectral Components}
The preceding calculations of reverberation time lags assume that the light curves of the directly observed continuum and reflected emission can be measured directly. In reality, however, light curves are obtained in spectral bands which are merely \textit{dominated} by either the continuum emission (the `hard' band at 1-4\keV) or the reflection from the accretion disc (the `soft' band at 0.3-1\keV). This means that the measured time lags will be `diluted' as the measured `reflection' will include promptly-arriving photons from the continuum and the `continuum' will include late-arriving photons from the reflection.

\begin{figure}
	\centering
	\includegraphics[height=85mm,angle=270]{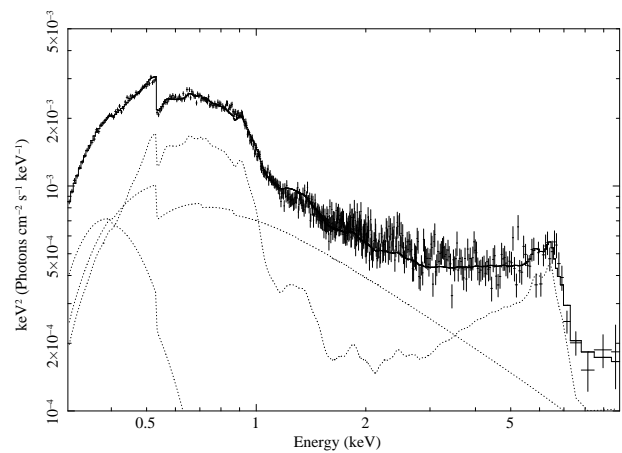}
	\caption{The time-averaged X-ray spectrum of 1H\,0707-495 observed using the EPIC pn detector on XMM Newton in October 2008, modelled as the sum of a direct power law component, reflection from an ionised accretion disc (using the \textsc{reflionx} model of \citealt{ross_fabian}) and black body emission from the disc.}
	\label{1h0707_eeuf.fig}
\end{figure}

Fig. \ref{1h0707_eeuf.fig} shows the X-ray spectrum of 1H\,0707-495 modelled as the sum of a direct power law component, reflection from the accretion disc and black body emission from the disc. Integrating the number of photons detected from each of the components over each energy band reveals that 0.3-1\keV\ band (dominated by the reflection component) contains a contribution from the primary continuum at 70 per cent of the number of photons detected in this band from the reflection. Likewise, the 1-4\keV\ band (dominated by the primary continuum) contains a contribution from the reflected emission at 60 per cent of the number of photons detected in this band from the power law. We assume the flux in the thermal, black body emission to be constant and therefore not to contribute to the lag.

\begin{table}
\centering
\caption{Contributions from the spectral components of 1H\,0707-495 from the January 2008 XMM Newton observation in the hard and soft bands from which reverberation time lags are calculated. Photon counts are normalised to the component said to be dominating in that band.}
\begin{tabular}{lcc}
  	\hline
   	\textbf{Component} & \textbf{0.3-1\thinspace keV} & \textbf{1-4\thinspace keV} \\
	\hline
	Power law continuum & 0.7 & 1 \\
	Reflection & 1 & 0.6 \\
	Black body & 0.4 & $10^{-4}$ \\	
	\hline
\end{tabular}
\label{1h0707_components.tab}
\end{table}

\begin{figure}
	\centering
	\includegraphics[width=85mm]{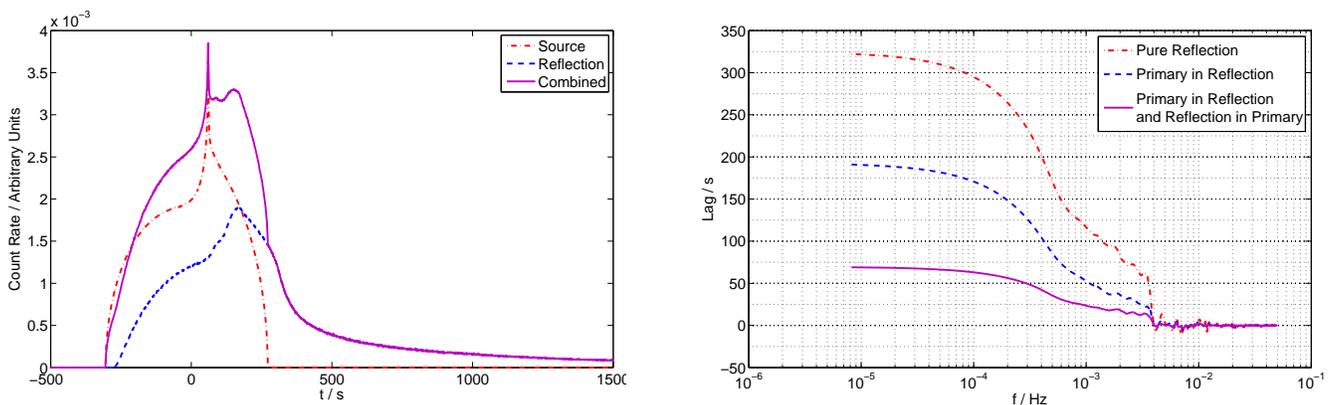}
	\caption{The combination of the transfer functions for reflection from the accretion disc and direct emission from the X-ray source, in the ratio found in 1H\,0707-495, to give the overall transfer function for the soft (0.3-1.0\keV) band.}
	\label{trf_comb.fig}
\end{figure}

To determine the effect of this dilution on the observed lag spectrum, a composite transfer function for the soft `reflection' band was formed by summing the transfer functions due to reflection from the accretion disc and for the direct emission seen from the primary X-ray source, weighted such that the integrated photon counts in the constituents are split in the ratio observed in the spectrum (Fig. \ref{trf_comb.fig}). Likewise, the transfer function in the hard `primary' band is constructed as the sum of contributions from the direct continuum emission and reflection from the accretion disc. These combined transfer functions are used to obtain the light curves that would be observed in the hard and soft bands with Equations \ref{hconv.equ} and \ref{sconv.equ}. Using the photon fluxes calculated above, the composite transfer functions are
\begin{align*}
	T_{0.3-1.0\mathrm{keV}} &= T_\mathrm{reflection} + 0.7T_\mathrm{source} \\
	T_{1.0-4.0\mathrm{keV}} &= T_\mathrm{source} + 0.6T_\mathrm{reflection}
\end{align*}

\begin{figure}
	\centering
	\includegraphics[width=85mm]{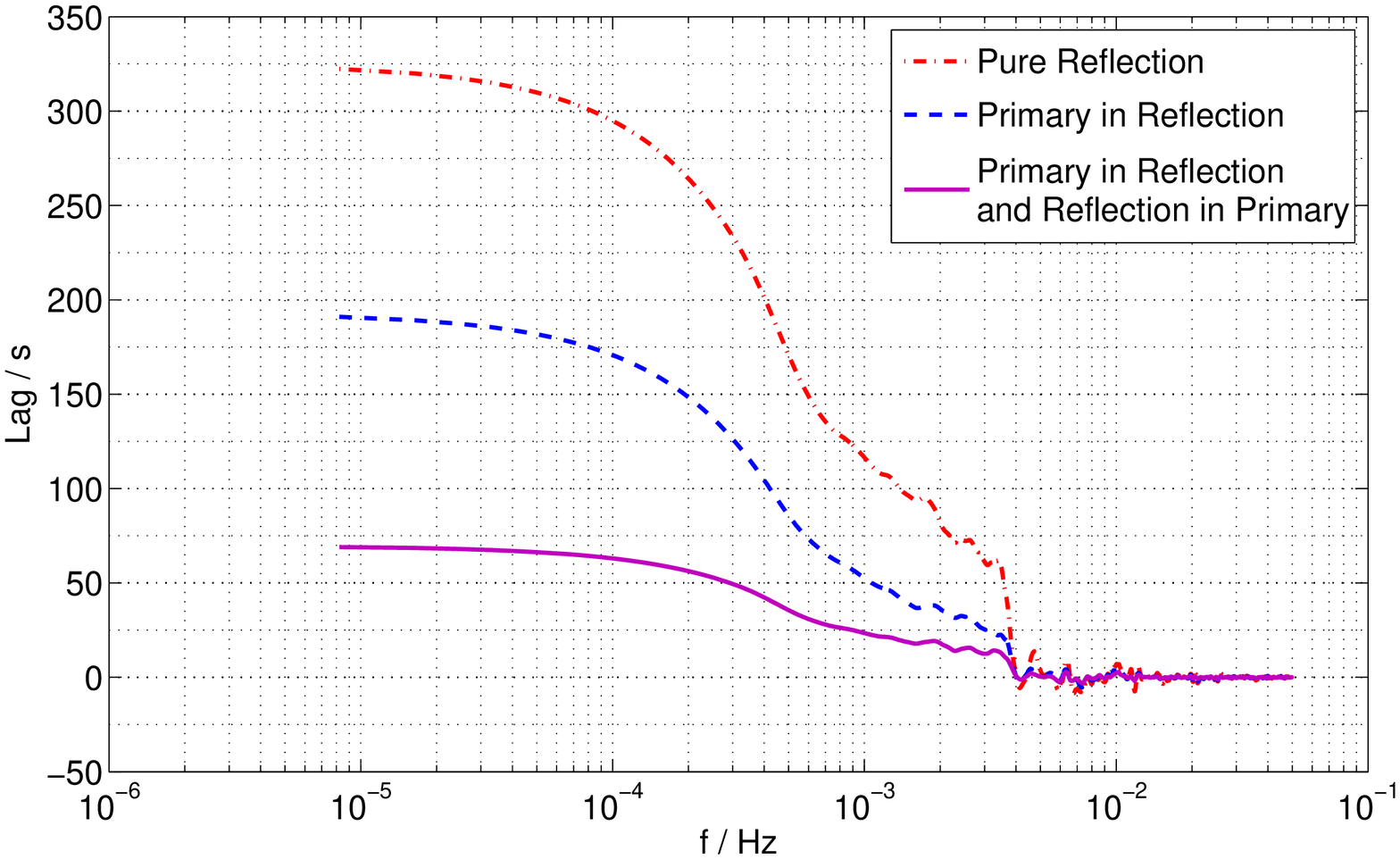}
	\caption{The effect of dilution of spectral components in the `primary' and `reflection' energy bands on the lag spectrum is illustrated by comparing the lag spectrum assuming a pure reflection component, the lag spectrum obtained using a reflection transfer function combined with a transfer function of the primary X-ray source contributing to this energy band and a lag spectrum also accounting for the reflection component contributing to the `primary' energy band in the calculation of the observed arrival time of the direct emission. In each case, the source is taken to extend radially out to 35\rg\ and vertically between 2.0 and 2.5\rg.}
	\label{lagspec_dilution.fig}
\end{figure}

Fig. \ref{lagspec_dilution.fig} shows the effect of dilution of the spectral components in the `primary' and `reflection' energy bands on the lag spectrum. It is clear that when the contribution from the primary X-ray source is included in the `reflection' band, the observed time lag is reduced by around 40 per cent compared to the `actual' time lag that would be seen if it were possible to directly observe the primary and reflected light curves. When, in addition to this, the contribution of the reflected component is included in the primary continuum band, the average time lag is further reduced. The observed time lag at $10^{-3}\thinspace\mathrm{Hz}$ can be reduced by up to 75 per cent from the `actual' time lag.

It is evident that when interpreting the observed time lags between the primary continuum emission and reflection in terms of the light travel time to constrain the location of the X-ray source with respect to the reflector, it is important to interpret the observed `hard' and `soft' spectral bands as combinations of the primary and reflected emission, rather than simply using them as proxies for the respective components.

\section{Propagation Effects}
\label{prop.sec}
When considering X-ray sources spanning radially out to around 35\rg, it is not physically meaningful to consider the source varying instantaneously through its entire extent, as has been assumed to this point where, in evaluating the transfer functions, all of the rays were taken to originate from the X-ray source at the same time. A fluctuation in the intensity of the primary X-ray source will propagate throughout its extent and in fact must do so in a time greater than the light travel time across the source region (which would be 35\rg$/ c$ for a source spanning 35\rg).

There are a number of possible scenarios for the propagation of fluctuations in the intensity of the primary X-ray source, which can arise due to changes in the density or energy/temperature of the particles in the corona. It is possible that the fluctuations in the corona originate from the injection of energy by a process close to the central black hole and that fluctuations in the X-rays emitted from the corona originate near the centre and propagate outwards. Alternatively, it is possible that energy is injected into the corona from the surface of the accretion disc, either directly or through magnetic fields that are anchored to the accretion disc that accelerate the particles of the corona to high energies. In this case, one might expect that fluctuations arise from fluctuations in the accretion disc (for instance, variations in the mass accretion rate or in the magnetic flux density as field lines are `accreted' inwards through the disc as per \citealt{beckwith+09} or \citealt{mckinney+2012}). Such fluctuations are likely to propagate inwards through the accretion disc and will cause fluctuations in the corona that originate at the edge and move inwards. Finally, it could be that there is no ordered propagation of fluctuations through the corona, rather variations in the intensity of the primary X-ray source are stochastic in nature, occurring randomly throughout the source region (for example, they arise due to isolated magnetic flaring events with no change in the bulk properties of the corona over the time-scale of the observations).

We investigate whether it is possible to discriminate between these scenarios in the observed lag spectra by computing transfer functions for the primary X-ray source and the reflection from the accretion disc in which the rays originate from the source at variable times. The influence of propagation effects is investigated on the lag spectrum of an X-ray emitting region extending radially outward to 35\rg\ and between 2 and 2.5\rg\ above the plane of the accretion disc and the full effects of dilution of the spectral components in the two energy bands are considered with the dilution fractions as found in 1H\,0707-495.

Random fluctuations in the X-ray source (if the changes in the observed X-ray luminosity originate in random flaring events throughout the source region whose net effect is to change the total count rate we observe; the variability arising probabilistically) are simulated by assigning a random start time to the rays up to a maximum value (here taken to be a multiple of the light travel time from the centre to the edge of the source region) which defines a characteristic timescale over which variations occur throughout the source region. These rays make up the transfer function, so in this model, constructed to give insight into how such a system would manifest itself in reverberation measurements, the random fluctuations are not the cause of the variability, rather given the observed variability, they dictate its spatial propagation through the source region and therefore how it is passed into the X-ray reflection.

\begin{figure}
	\centering
	\includegraphics[width=85mm]{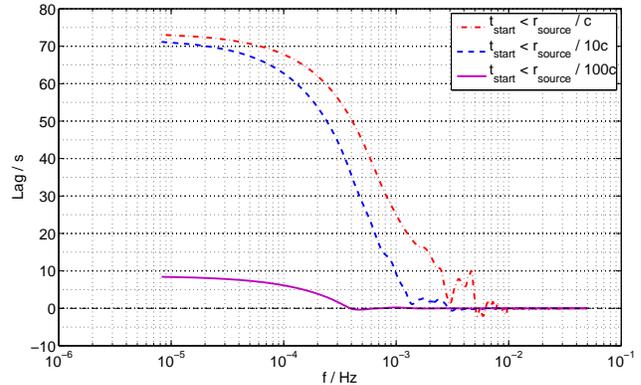}
	\caption{The lag spectrum between the hard (1.0-4.0\keV, dominated by the primary continuum) and soft (0.3-1.0\keV, dominated by reflection from the acctrion disc) X-ray bands, taking into account the contributions of both spectral components to each band, allowing the rays from the primary X-ray source to have random start times up to a maximum (as a function of the radius of the source). The emitting region is located between 2.0 and 2.5\rg\ above the accretion disc and extends radially out to 35\rg.}
	\label{lagspec_randstart.fig}
\end{figure}

The lag spectrum arising from such a scenario, with increasing ranges in the ray start times in each transfer function is shown in Fig. \ref{lagspec_randstart.fig}. Randomising the start times of rays through the source region results in the dilution of the measured lag between the arrival of the primary and reflected X-rays, with the lag between the primary and reflected components tending to zero when the timescale of the random flaring events is much greater than the light travel time between the X-ray source and the reflector.

To simulate the effect of fluctuations in the source luminosity propagating outwards from the centre of the source region, the start time of a ray is taken to be proportional to its originating radius within the region, where the constant of proportionality represents the speed at which fluctuations move radially through the source region, corresponding to dynamical timescales within the corona, and is the same for all rays. Likewise, the effect of fluctuations propagating inwards from the edge of the X-ray source is simulated by setting the start time of the rays proportional to their initial radial distance from the edge of the source region.

\begin{figure*}
\centering
\begin{minipage}{170mm}
\subfigure[Inward propagation]{
\includegraphics[width=85mm]{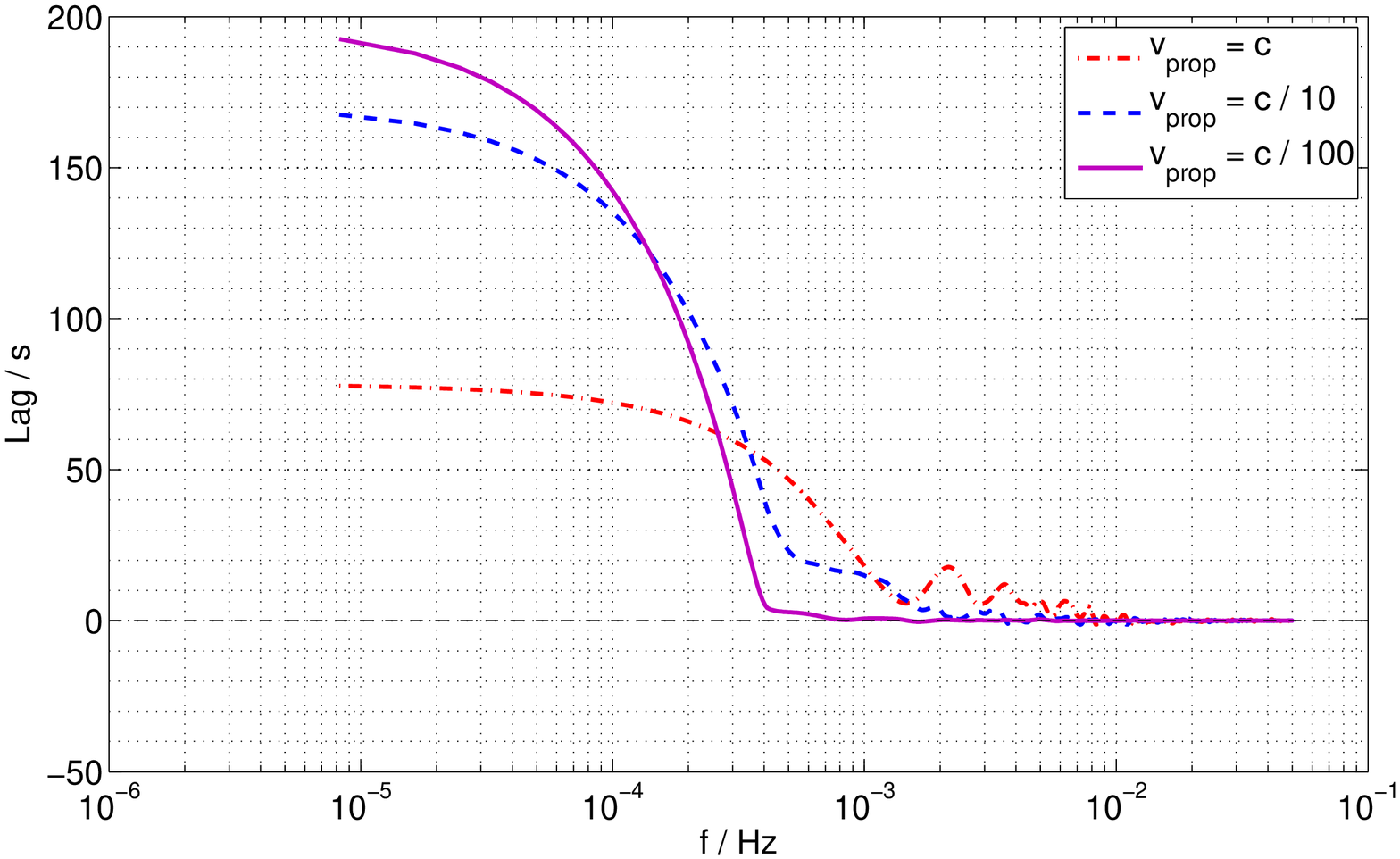}
\label{lagspec_prop.fig:in}
}
\subfigure[Outward propagation]{
\includegraphics[width=85mm]{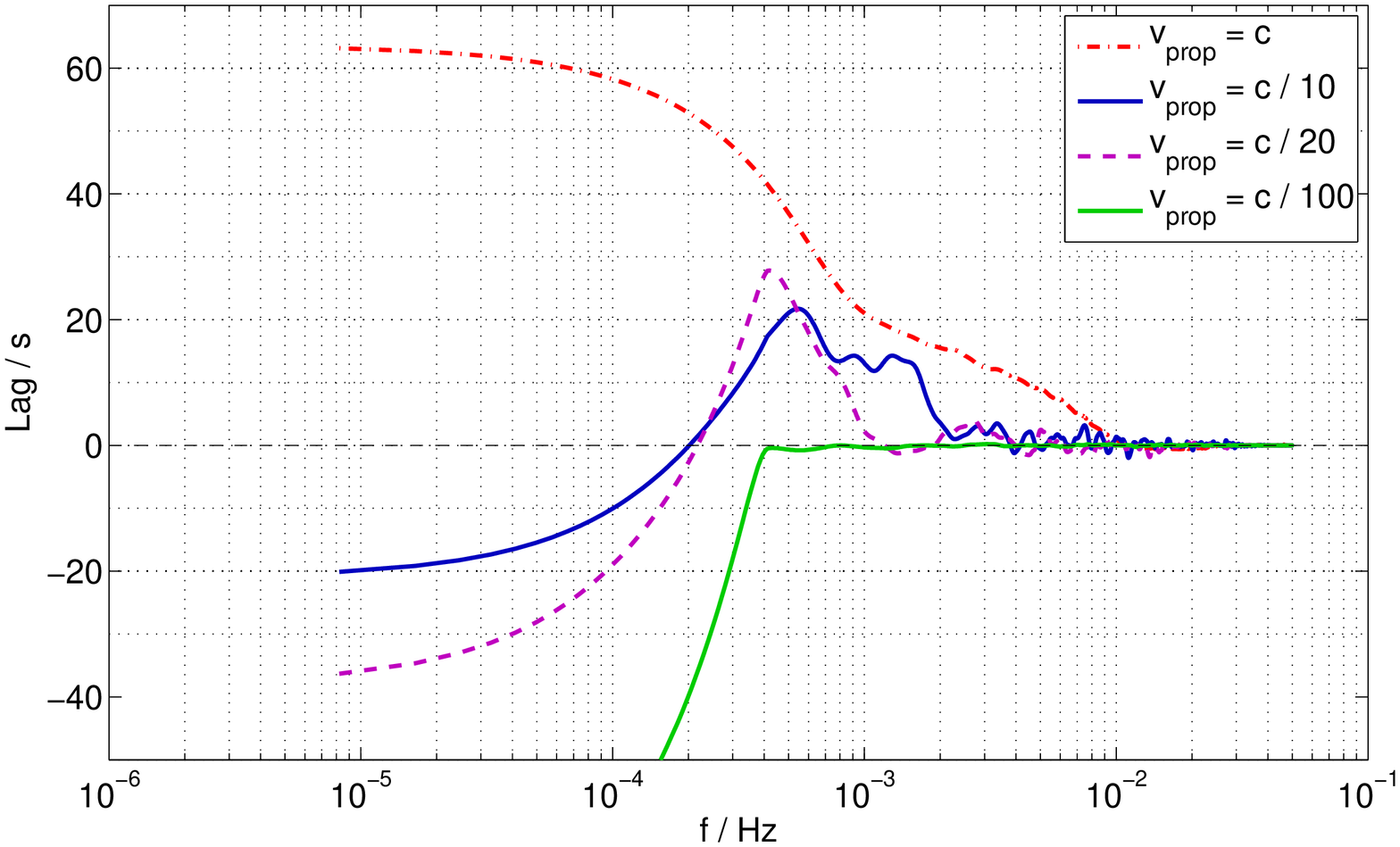}
\label{lagspec_prop.fig:out}
}
\caption[]{Lag spectra arising from fluctuations in the X-ray source luminosity propagating \subref{lagspec_prop.fig:in} inward from the edge of the defined source region and \subref{lagspec_prop.fig:out} outward from the centre. In each case, the lag spectra are shown for varying velocities at which the fluctuations propagate radially through the source region. The lag spectra are between the hard (1.0-4.0\keV, dominated by the primary continuum) and soft (0.3-1.0\keV, dominated by reflection from the acctrion disc) X-ray bands, taking into account the contributions of both spectral components to each band and the emitting region is located between 2.0 and 2.5\rg\ above the accretion disc and extends radially out to 35\rg.}
\label{lagspec_prop.fig}
\end{minipage}
\end{figure*}

The propagation of luminosity fluctuations inward from the edge of a defined source region delays the arrival time of the reflected radiation with respect to the primary continuum. The resulting lag spectrum is shown in Fig. \ref{lagspec_prop.fig:in}. As the speed at which fluctuations propagate through the source region is decreased, the measured lag increases and the lag at high frequencies is damped to zero over a broader frequency range.

Conversely, Fig. \ref{lagspec_prop.fig:out} shows theoretical lag spectra in which the fluctuation in source intensity propagates outward from the centre at varying speeds. While the propagation of fluctuations outward throughout the source region reduces the apparent time lag between the primary continuum and reflection, the most notable effect is the turnover at low frequencies, $< 10^{-4}\thinspace\mathrm{Hz}$, once the fluctuations propagate slower than approximately one tenth of the speed of light. In these situations, low frequency variations appear in the soft X-ray band, representing the `reflection' before the hard band, representing the `primary continuum.'

Due to the propagation delay, X-rays from the outer parts of the source, which due to their larger area make up the majority of the flux received from the primary X-ray source, are delayed in their emission and thus their arrival at the observer. On the other hand, X-rays emitted from parts of the source closer to the black hole are more likely to be reflected from the accretion disc rather than observed directly in the continuum as rays passing closer to the black hole will be focussed on to the accretion disc by the strong gravity. This means that, comparatively, the reflected X-rays originate from parts of the source region closer to the black hole than the observed continuum and since the propagation time of the fluctuation is greater than the light travel time to the reflector, the average arrival time of the soft X-ray band at the observer is slightly earlier than the average arrival time of the hard band once the full contributions of the continuum and reflection in each band are accounted for (the dilution), resulting in the soft band slightly leading the hard band at the lowest variability frequencies (this effect is phase-wrapped out of the variability at higher frequencies where the soft band is seen to lag behind the hard band).

If the speed at which the fluctuations propagate is further reduced, the low frequency variability in the observed hard X-ray band (dominated by the primary continuum) lags further behind that in the soft band while the high frequency part of the lag spectrum decays as the increasing lags cause these components to be phase-wrapped out.

\section{The Lag Spectrum of 1H\,0707-495}
The lag spectrum obtained for the narrow line Seyfert 1 galaxy, 1H\,0707-495 \citep[see][ detailing the data reduction in the context of computing the lag spectrum]{zoghbi+09} between the 1.0-4.0\keV\ band (dominated by the power law continuum) 0.3-1.0\keV\ band (dominated by reflection from the accretion disc) is compared to the theoretical lag spectra computed from the reflection of an X-ray continuum arising from an extended source from the accretion disc.

It is clear that the high frequency part of the lag spectrum ($> 10^{-3}\thinspace\mathrm{Hz}$) in which the soft `reflection' band lags behind the hard `primary' band by around 30\s\ is explained naturally by an X-ray source extending outward from the rotation axis to 35\rg\ and located around 2\rg\ above the plane of the accretion disc which, once the effects of dilution are taken into account matches the lag of 30\s\ at $10^{-3}$\Hz\ and the decay of the lag towards higher frequencies (solid line in Fig. \ref{lagspec_dilution.fig}). This is consistent with the extent of the primary X-ray source derived by \citet{wilkins_fabian_2012} from the observed emissivity profile of the reflected iron K$\alpha$ line from the accretion disc, showing that the two independent techniques compliment one another in placing constraints on the properties of the source.

It is assumed here that the X-ray source has a constant luminosity throughout its extent. In reality, it is likely that the X-ray emission will vary in luminosity based upon its location, with, perhaps, the emission being concentrated towards the centre of the source region. Assuming constant source luminosity, however, minimised the number of free parameters in this model and finding the X-ray source under this assumption has allowed the data itself to constrain the location of the bulk of the X-ray emission. This result is the vital first step in more detailed modelling of the structure of the corona which must simultaneously satisfy the constraints from the observed lag spectrum, emissivity profile and energy spectrum.

The simple model does not, however, explain the lag spectrum at low frequencies where the hard band is seen to lag behind the soft band. Such a turnover in the lag spectrum may be, in part, accounted for by the propagation of luminosity fluctuations through the primary X-ray emitting region (in fact, fluctuations must propagate in a finite time through a spatially extended source). If fluctuations in the X-ray source luminosity propagate outward from the centre of the source region at velocities of order one tenth of the speed of light, the low frequency variations in the soft band lead those in the hard band, while the high frequency variability shows the reverberation lag with the soft band lagging behind the hard, shown in Fig. \ref{lagspec_prop.fig:out}.

The hard X-ray band lagging behind the soft band at low frequencies may suggest a scenario in which energy is injected into the corona at the centre, close to the black hole, then propagates outwards. It should be noted that this model does not make any assumptions about the origin of the fluctuations in the X-ray source luminosity or their power spectral density (PSD). In fact, since the theoretical lag spectra have been computed by convolving the observed hard band light curve of 1H\,0707-495 with the transfer functions corresponding to the X-ray source and the reflection, the observed PSD is accounted for inherently in the model. Neither are any assumptions made of the underlying energy transport mechanism between the accretion flow into the X-ray emitting region. Rather, we account for the observed lag spectrum of AGN in terms of the propagation of variations at a characteristic velocity (analogous to a dynamical timescale) through an X-ray emitting region which is also constrained by independent, spectral measurements of the accretion disc emissivity profile, and the response of the reflecting accretion disc.

In January 2011, 1H\,0707-495 was observed to fall into a low flux state in which the spectrum was almost entirely dominated by reflection from the accretion disc with almost no flux observed from the power law continuum \citep{1h0707_jan11}. The reflection spectrum from the accretion disc was identified with a very steeply falling emissivity profile implying a small source concentrated close to the black hole. This can be reconciled with the energy that accelerates particles in the X-ray source originating from the central regions and then propagating outwards to form an extended source region. When the available energy decreases, particles are still accelerated close to the black hole, though since less energy is available, it is not propagated outwards meaning the source is unable to extend, reducing the detected continuum emission as the emitting volume is smaller and the emission closer to the black hole tends to be focussed on to the accretion disc to be detected after reflection.

It is possible that the fluctuations in the luminosity of the primary X-ray source are due to fluctuations into the mass accretion rate onto the central black hole (varying the energy that is released during the gravitational infall), with the fluctuations in accretion rate resulting in density fluctuations spiralling inwards through the accretion disc. Such a model is proposed by \citet{kotov+2001} and \citet{arevalo+2006} in which the X-ray variability of Galactic black hole binaries is explained by fluctuations in the accretion disc propagating inwards, modulating the luminosity of the X-ray emitting region as they do so. This model  does not account for how the accretion disc is coupled to the corona and only finds time lags in different energy bands of the emission from the primary X-ray source with measurements in energy bands both dominated by the continuum. In the case of AGN, however, the dominant spectral component in the soft band is the reflection from the accretion disc with a characteristic lag spectrum resulting from reverberation of variability in the source.

The simple reverberation from a fluctuation that propagates outwards through the X-ray source does not, however, reproduce the entirety of the hard lag which is observed to be longer. Other physical effects are likely to contribute to this. As well as fluctuations propagating through the source, the source properties are likely to vary throughout its extent, such as its density and energy per particle, meaning that the luminosity and spectrum (characterised by the power law slope) of X-rays emitted from different parts of the source may change. A greater Compton optical depth, up to unity, will result in the seed photons undergoing more scatterings and therefore reaching higher energies, producing a harder continuum spectrum. However, increasing the optical depth above unity will mean the photons undergo so many scatterings, the net effect is that energy is distributed more between the photons, resulting in a softer spectrum with more photons at lower energies. Likewise, increasing the average energy of the particles in the corona will produce a harder spectrum where more energy is transferred to the photons \citep[see, \textit{e.g.},][]{titarchuk-94}.

The observed hard lag at low frequencies can result from the X-ray source itself if the fluctuations reach parts of the source producing a harder spectrum after they reach parts of the source producing a soft spectrum. This will mean at low frequencies, variability in the hard X-ray band will lag behind that in the soft band. If this effect is to add to the hard lag produced intrinsically by the propagation of fluctuations through the source, the central regions of the X-ray source produce a softer power law continuum spectrum than the outer parts. This is possible if the average energy of the particles in the corona is greater in the outer parts. If it is only the higher energy particles that are able to travel out further from the central part into which the energy is injected from the accretion flow with the lower energy particles remaining in the centre, the average energy of particles in the outer parts of the source region will be greater, producing a harder spectrum here. Alternatively, a softer power law continuum could be produced by the innermost parts of the source if the density is much greater here such that the optical depth is greater than unity (such that the increasing density produces a softer spectrum), though it is unlikely a Compton-thick source extends further out than the innermost stable orbit of the accretion disc as this corona would obscure the reflected emission seen from the disc.

Finally, it is possible that the fluctuations in source luminosity propagate inwards from the edge of the source region and the inner parts of the X-ray source produce a harder power law continuum as the Compton depth and particle energy increases towards the centre, however for this to be the case, the variation in spectral slope but be sufficient across the source that the intrinsic time lag between the energy bands within the source is able to counteract the increased soft lag (the soft, reflection band lagging behind the hard, continuum band) that results from the fluctuations propagating inwards.

The applicability to 1H\,0707-495 of a simple model consisting of two power law continua with different slopes with the hard power law lagging behind the soft reproducing the hard lag is considered by \citet{kara+12}. Detailed modelling of the lag spectra arising from such a scenario would require detailed analysis of the spectral contributions that could come from power law continua with different slopes and will be considered more fully in future work.

\section{A Second Example: IRAS\,13224-3809}
\citet{iras} obtained the lag spectrum of the narrow line Seyfert 1 galaxy IRAS\,13224-3809 from a long observation of the X-ray emission from this source with XMM Newton. The lag spectrum is of the same form is that of 1H\,0707-495 with a reverberation lag (the soft band corresponding to reflection from the accretion disc lagging behind the hard band corresponding to the directly observed continuum) of around 100\s\ at a frequency of around $4\times 10^{-3}\,\mathrm{Hz}$ and the turnover at low frequencies with the hard band lagging behind the soft.

The longer observed lag (once dilution has been taken into account) and the shift of the form of the lag spectrum to lower frequencies is consistent with the greater mass of the central black hole in IRAS\,13224-3809 than in 1H\,0707-495 (around $10^7$\Msun) and with the X-ray source being more compact. The emissivity profile of the accretion disc implies the X-ray source is contained within the innermost 10\rg, which, from Fig. \ref{lags_radius.fig:spec}, implies in itself a shift of the reverberation lag to lower frequencies and a longer reverberation lag than for the more extended source in 1H\,0707-495 since there are fewer possible ray paths from an extended source and the rays pass closer to the central black hole, increasing the influence of the Shapiro delay.

\section{Conclusions}
In order to interpret observed lag spectra and reverberation lags between the primary continuum and reflected X-ray emission in AGN, theoretical lag spectra were calculated for reflection of the X-ray continuum from the accretion disc by computing transfer functions in general relativistic ray tracing simulations which were convolved with the observed hard band X-ray light curve of 1H\,0707-495. Lag spectra were computed for isotropic point sources at varying heights as well as a variety of spatially extended sources.

The Shapiro delay of light travelling close to the central black hole as a significant effect on the travel time of rays from the source to the reflector and to the observer. Therefore, when inferring the location and properties of the primary X-ray source from measurements of reverberation lags, it is important to account for general relativistic effects rather than directly equating the measured lag to a light travel time and the equivalent distance in flat, Euclidean space.

When interpreting observed reverberation lags, it is necessary to consider the contribution of all spectral components to the `hard' and `soft' bands rather than taking them to be direct proxies for the primary and reflected radiation. The `hard' and `soft' X-ray bands are seen to have significant contributions from both the primary continuum and reflected spectral components in AGN such as 1H\,0707-495 which causes the observed time lag between the arrival of the primary continuum and reflection components to be diluted, reducing the observed lag by up to 75 per cent compared to the `intrinsic' light travel time delay. It is therefore vital to have adequate spectral models to understand the contributions to the energy bands when interpreting reverberation lags.

By comparing the observed lag spectrum of 1H\,0707-495, we are able to constrain the primary X-ray continuum to originating from a source region extending radially outwards to around 35\rg\ from the rotation axis, located around 2\rg\ above the plane of the accretion disc. This is consistent with constraints on the location and extent of the X-ray source derived independently from the observed X-ray reflection emissivity profile of the accretion disc.

The propagation of luminosity fluctuations through an extended X-ray source is considered, since it is unphysical to think of the luminosity of a spatially extended source varying instantaneously. It is demonstrated that if the fluctuations in luminosity are allowed to propagate radially outward from the centre of the source at a velocity (corresponding to a dynamical timescale in the corona) of order a tenth of the speed of light, the resulting lag spectrum turns over at low frequencies, $< 10^{-4}\thinspace\mathrm{Hz}$ and variability in the soft band (dominated by reflection) leads that in the hard band (dominated by the primary continuum) as is seen in AGN such as 1H\,0707-495. This suggests that the hard lag could arise, in part, from a scenario in which energy is injected into the corona from the accretion flow very close to the central black hole and then propagates outwards into the spatially extended source region.

In this work, no assumptions are made as to the underlying physical processes through which energy is released from the accretion flow into the X-ray emitting corona, rather constraints are placed upon the properties the X-ray source must have to be consistent with observations in the context of the reverberation off an accretion disc surrounding a black hole in general relativity. These findings may be compared to more physical models of coron\ae\ or indeed rays may be traced in the same framework from X-ray sources produced by these models to constrain the underlying physical processes. It is argued that the energetics of the X-ray emitting corona must be dominated by magnetic fields and corona composed of magnetic flux loops emerging from the accretion disc due tot he magneto-rotational instability have been studied extensively by, among others, \citet{uzdensky+08}. They describe the corona as a statistical ensemble of magnetic flux loops in (magnetic) pressure equilibrium with their surroundings, allowing them to undergo reconnection and predict that the magnetic energy density should be concentrated at low heights above the accretion disc (within a few times the scale height of the disc). Assuming a thin accretion disc with aspect ratio $H/r \sim 0.1$, this is roughly consistent with our findings from the reverberation lags that the bulk of the X-ray source is located within a few gravitational radii above the accretion disc, though more detailed theoretical work tailored to these observations would be required to place any firm constraints on the underlying process powering the corona.

\section*{Acknowledgements}
Thanks must go to Erin Kara, Ed Cackett and Chris Reynolds for enlightening and extremely useful discussions during this work.

\bibliographystyle{mnras}
\bibliography{agn}

\begin{thebibliography}{}

\bibitem[\protect\citeauthoryear{{Ar{\'e}valo} \& {Uttley}}{{Ar{\'e}valo} \&
  {Uttley}}{2006}]{arevalo+2006}
{Ar{\'e}valo} P.,  {Uttley} P., 2006, \mnras, 367, 801

\bibitem[\protect\citeauthoryear{{Beckwith}, {Hawley}, \& {Krolik}}{{Beckwith}
  et~al.}{2009}]{beckwith+09}
{Beckwith} K., {Hawley} J.~F.,  {Krolik} J.~H., 2009, \apj, 707, 428

\bibitem[\protect\citeauthoryear{{de Marco} et~al.}{{de Marco}
  et~al.}{2012}]{demarco+2012}
{de Marco} B., {Ponti} G., {Cappi} M., {Dadina} M., {Uttley} P., {Cackett}
  E.~M., {Fabian} A.~C.,  {Miniutti} G., 2012, ArXiv e-prints

\bibitem[\protect\citeauthoryear{{de Marco} et~al.}{{de Marco}
  et~al.}{2011}]{demarco+2011}
{de Marco} B., {Ponti} G., {Uttley} P., {Cappi} M., {Dadina} M., {Fabian}
  A.~C.,  {Miniutti} G., 2011, \mnras, 417, L98

\bibitem[\protect\citeauthoryear{{Emmanoulopoulos}, {McHardy}, \&
  {Papadakis}}{{Emmanoulopoulos} et~al.}{2011}]{emmanoul+2011}
{Emmanoulopoulos} D., {McHardy} I.~M.,  {Papadakis} I.~E., 2011, \mnras, 416,
  L94

\bibitem[\protect\citeauthoryear{Fabian et~al.}{Fabian et~al.}{2012}]{iras}
Fabian A.~C. et~al., 2012, arXiv e-Prints

\bibitem[\protect\citeauthoryear{{Fabian} et~al.}{{Fabian}
  et~al.}{1989}]{fabian+89}
{Fabian} A.~C., {Rees} M.~J., {Stella} L.,  {White} N.~E., 1989, \mnras, 238,
  729

\bibitem[\protect\citeauthoryear{{Fabian} et~al.}{{Fabian}
  et~al.}{2009}]{fabian+09}
{Fabian} A.~C. et~al., 2009, \nat, 459, 540

\bibitem[\protect\citeauthoryear{{Fabian} et~al.}{{Fabian}
  et~al.}{2011}]{1h0707_jan11}
{Fabian} A.~C. et~al., 2011, \mnras, 1970

\bibitem[\protect\citeauthoryear{{George} \& {Fabian}}{{George} \&
  {Fabian}}{1991}]{george_fabian}
{George} I.~M.,  {Fabian} A.~C., 1991, \mnras, 249, 352

\bibitem[\protect\citeauthoryear{{Gilfanov}, {Churazov}, \&
  {Revnivtsev}}{{Gilfanov} et~al.}{2000}]{gilfanov+2000}
{Gilfanov} M., {Churazov} E.,  {Revnivtsev} M., 2000, \mnras, 316, 923

\bibitem[\protect\citeauthoryear{{Kara} et~al.}{{Kara} et~al.}{2012}]{kara+12}
{Kara} E., {Fabian} A.~C., {Cackett} E.~M., {Steiner} J., {Uttley} P.,
  {Wilkins} D.~R.,  {Zoghbi} A., 2012, MNRAS submitted

\bibitem[\protect\citeauthoryear{{Kotov}, {Churazov}, \& {Gilfanov}}{{Kotov}
  et~al.}{2001}]{kotov+2001}
{Kotov} O., {Churazov} E.,  {Gilfanov} M., 2001, \mnras, 327, 799

\bibitem[\protect\citeauthoryear{{McKinney}, {Tchekhovskoy}, \&
  {Blandford}}{{McKinney} et~al.}{2012}]{mckinney+2012}
{McKinney} J.~C., {Tchekhovskoy} A.,  {Blandford} R.~D., 2012, \mnras, 423,
  3083

\bibitem[\protect\citeauthoryear{{Nowak} et~al.}{{Nowak}
  et~al.}{1999}]{nowak+99}
{Nowak} M.~A., {Vaughan} B.~A., {Wilms} J., {Dove} J.~B.,  {Begelman} M.~C.,
  1999, \apj, 510, 874

\bibitem[\protect\citeauthoryear{{Reynolds} et~al.}{{Reynolds}
  et~al.}{1999}]{reynolds+99}
{Reynolds} C.~S., {Young} A.~J., {Begelman} M.~C.,  {Fabian} A.~C., 1999, \apj,
  514, 164

\bibitem[\protect\citeauthoryear{{Ross} \& {Fabian}}{{Ross} \&
  {Fabian}}{2005}]{ross_fabian}
{Ross} R.~R.,  {Fabian} A.~C., 2005, \mnras, 358, 211

\bibitem[\protect\citeauthoryear{{Shakura} \& {Sunyaev}}{{Shakura} \&
  {Sunyaev}}{1973}]{shaksun}
{Shakura} N.~I.,  {Sunyaev} R.~A., 1973, \aap, 24, 337

\bibitem[\protect\citeauthoryear{Shapiro}{Shapiro}{1964}]{shapiro}
Shapiro I.~I., 1964, Phys. Rev. Lett., 13, 789

\bibitem[\protect\citeauthoryear{{Stella}}{{Stella}}{1990}]{stella-90}
{Stella} L., 1990, \nat, 344, 747

\bibitem[\protect\citeauthoryear{{Titarchuk}}{{Titarchuk}}{1994}]{titarchuk-94}
{Titarchuk} L., 1994, \apj, 434, 570

\bibitem[\protect\citeauthoryear{{Uttley} et~al.}{{Uttley}
  et~al.}{2011}]{uttley+2011}
{Uttley} P., {Wilkinson} T., {Cassatella} P., {Wilms} J., {Pottschmidt} K.,
  {Hanke} M.,  {B{\"o}ck} M., 2011, \mnras, 414, L60

\bibitem[\protect\citeauthoryear{{Uzdensky} \& {Goodman}}{{Uzdensky} \&
  {Goodman}}{2008}]{uzdensky+08}
{Uzdensky} D.~A.,  {Goodman} J., 2008, \apj, 682, 608

\bibitem[\protect\citeauthoryear{{Wilkins} \& {Fabian}}{{Wilkins} \&
  {Fabian}}{2011}]{wilkins_fabian_2011a}
{Wilkins} D.~R.,  {Fabian} A.~C., 2011, \mnras, 414, 1269

\bibitem[\protect\citeauthoryear{{Wilkins} \& {Fabian}}{{Wilkins} \&
  {Fabian}}{2012}]{wilkins_fabian_2012}
{Wilkins} D.~R.,  {Fabian} A.~C., 2012, \mnras, 424, 1284

\bibitem[\protect\citeauthoryear{{Young} \& {Reynolds}}{{Young} \&
  {Reynolds}}{2000}]{young_reynolds}
{Young} A.~J.,  {Reynolds} C.~S., 2000, \apj, 529, 101

\bibitem[\protect\citeauthoryear{{Zhou} \& {Wang}}{{Zhou} \&
  {Wang}}{2005}]{zhou_wang}
{Zhou} X.-L.,  {Wang} J.-M., 2005, \apjl, 618, L83

\bibitem[\protect\citeauthoryear{{Zoghbi} \& {Fabian}}{{Zoghbi} \&
  {Fabian}}{2011}]{zoghbi+2011}
{Zoghbi} A.,  {Fabian} A.~C., 2011, \mnras, 418, 2642

\bibitem[\protect\citeauthoryear{{Zoghbi} et~al.}{{Zoghbi}
  et~al.}{2012}]{zoghbi+2012}
{Zoghbi} A., {Fabian} A.~C., {Reynolds} C.~S.,  {Cackett} E.~M., 2012, \mnras,
  422, 129

\bibitem[\protect\citeauthoryear{{Zoghbi} et~al.}{{Zoghbi}
  et~al.}{2010}]{zoghbi+09}
{Zoghbi} A., {Fabian} A.~C., {Uttley} P., {Miniutti} G., {Gallo} L.~C.,
  {Reynolds} C.~S., {Miller} J.~M.,  {Ponti} G., 2010, \mnras, 401, 2419

\end{thebibliography}

\label{lastpage}

\end{document}